\magnification=\magstep1

\baselineskip=12pt plus 1pt 
\parindent=18pt
\parskip=0pt

\font\bigbold=cmbx10 scaled \magstep2
\font\bigslant=cmsl10 scaled \magstep1

\footline={\ifnum\pageno>0
           \hfil--\thinspace\folio\kern .2em--\hfil
           \else\hfil\fi}

\def\section#1{\bigbreak\leftline{\bf #1}\medskip\nobreak}

\def\prain{p_{\hbox{\sevenrm R}}}
\def\pnorain{p_{\hbox{\sevenrm R\kern-5.3pt}\raise.3pt\hbox{\sevenrm/}}}
\def\teff{t_{\rm eff}}

%less (greater) than or approximately equal
\def\lsim{
      \mathchoice
          {\mathrel{\raise.325ex\hbox{$\displaystyle<$}
                    \mkern-14mu
                    \lower.625ex\hbox{$\displaystyle\sim$}}}
          {\mathrel{\raise.325ex\hbox{$\textstyle<$}
                    \mkern-14mu
                    \lower.625ex\hbox{$\textstyle\sim$}}}
          {\mathrel{\raise.25ex\hbox{$\scriptstyle<$}
                    \mkern-13.5mu
                    \lower.55ex\hbox{$\scriptstyle\sim$}}}
          {\mathrel{\raise.2ex\hbox{$\scriptscriptstyle<$}
                    \mkern-13mu
                    \lower.5ex\hbox{$\scriptscriptstyle\sim$}}}
         }

\def\gsim{
      \mathchoice
          {\mathrel{\raise.325ex\hbox{$\displaystyle>$}
                    \mkern-14mu
                    \lower.625ex\hbox{$\displaystyle\sim$}}}
          {\mathrel{\raise.325ex\hbox{$\textstyle>$}
                    \mkern-14mu
                    \lower.625ex\hbox{$\textstyle\sim$}}}
          {\mathrel{\raise.25ex\hbox{$\scriptstyle>$}
                    \mkern-13.5mu
                    \lower.55ex\hbox{$\scriptstyle\sim$}}}
          {\mathrel{\raise.2ex\hbox{$\scriptscriptstyle>$}
                    \mkern-13mu
                    \lower.5ex\hbox{$\scriptscriptstyle\sim$}}}
         }

\newcount\refno
\refno=0
\def\ref{\advance\refno by 1\smallbreak\item{\the\refno.}}

\hyphenation{un-pre-dict-a-bil-i-ty}

\pageno=-1

\line{}
\vskip .75truein plus .15fil

\centerline{\bigbold Unpredictability, Information, and Chaos}

\vskip .4truein

\centerline{\bigslant Carlton M.~Caves\/$^{\hbox{\sevenrm (a)}}$ and 
R\"udiger Schack\/$^{\hbox{\sevenrm (b)}}$}

\vskip 0.25truein

\centerline{\hbox{}$^{\hbox{\sevenrm (a)}}$Center for Advanced Studies, 
Department of Physics and Astronomy,}
\centerline{University of New Mexico, Albuquerque, New Mexico \thinspace
87131-1156, USA}
\smallskip
\centerline{Telephone: (505)277-8674}
\centerline{FAX: (505)277-1520}
\centerline{E-mail: caves@tangelo.phys.unm.edu}

\vskip 0.25truein

\centerline{\hbox{}$^{\hbox{\sevenrm (b)}}$Department of Mathematics, 
Royal Holloway, University of London,}
\centerline{Egham, Surrey TW10 0EX, United Kingdom}
\smallskip
\centerline{E-mail: r.schack@rhbnc.ac.uk}

\vskip 0.4truein plus .25fil

\centerline{\sl ABSTRACT}

\vglue 0.25truein

A source of unpredictability is equivalent to a source of information:
unpredictability means not knowing which of a set of alternatives is
the actual one; determining the actual alternative yields information.
The degree of unpredictability is neatly quantified by the information
measure introduced by Shannon.  This perspective is applied to three
kinds of unpredictability in physics: the absolute unpredictability of
quantum mechanics, the unpredictability of the coarse-grained future
due to classical chaos, and the unpredictability of open systems.  The
incompatibility of the first two of these is the root of the difficulty
in defining quantum chaos, whereas the unpredictability of open systems,
it is suggested, can provide a unified characterization of chaos in 
classical and quantum dynamics.

\vskip 0pt plus .5fil

\centerline{Please direct all correspondence to C.~M. Caves at the 
above address.}
\centerline{22 pages; 1 figure; no tables}

\eject

\baselineskip=13pt plus 1pt minus 0.25pt
\pageno=1

\section{I. \ UNPREDICTABILITY AND INFORMATION}

Unpredictability.  To discuss it, we need a framework.  What is it?  How 
is it described, and---if we intend a scientific discussion---how is it 
quantified?  The goal in this introductory section is to suggest a 
framework for discussing and describing unpredictability.

Suppose one is interested in a particular set of alternatives.  
Unpredictability for these alternatives means simply that one can't 
say reliably which alternative is the actual one.  Several examples 
provide reference points for the discussion.

{\parindent=12pt\narrower
\item{$\bullet$}A horse race where a gambler bets on the winner.  The 
relevant alternatives for the gambler are the horses in the race.  

\item{$\bullet$}A game of frisbee golf played in a gusty wind.  The 
alternatives of interest to a participant are the possible paths of 
the frisbee---more precisely, histories of the frisbee's center-of-mass 
position and orientation as the frisbee is buffeted by the wind.

\item{$\bullet$}A New Mexico farmer growing chiles.  The chiles are 
infested by a pest that can be eliminated by applying a biological agent.  
The catch is that the agent is rendered ineffective if it rains within 
a day of application, not an unlikely event during the thunderstorm 
season of July and August.  The relevant alternatives for the farmer 
are whether or not it rains on a given day.

\item{$\bullet$}A photon incident on a polarizing beam splitter.  An 
experimenter, by adjusting the orientation of the beam splitter, selects
two orthogonal linear polarizations, which are sent along different 
paths by the beam splitter.  The alternatives of interest to the 
experimenter are these two orthogonal linear polarizations.
\smallbreak}

\noindent 
In these examples, there is an element of unpredictability, an 
uncertainty as to which alternative turns out to be the actual one.  
Though prediction is often thought of as having to do with future 
behavior---our examples have this flavor---in this article we do not 
include any temporal relationship in the notions of predictability and 
unpredictability.  {\it The key element in unpredictability is uncertainty 
about the actual alternative among a collection of alternatives}.  There
can be uncertainty about past as well as future events.  For example, 
the horse race might have been run last night; not knowing the outcome, 
two friends could bet on the winner today, facing the same uncertainty
about the outcome as they would have encountered before the race.

How is unpredictability described mathematically?  By assigning 
probabilities to the alternatives.  At the horse race the bettor 
assigns probabilities based on what he knows about the horses.  These 
probabilities are betting odds: if the bettor assigns a probability 
$1/n$ for a horse to win, it means that he is willing to place a bet 
on that horse at odds of $n$ to 1 or better.  The frisbee golfer, 
applying his experience to the present condition of the wind, assigns 
probabilities intuitively to various possible histories of the 
frisbee's motion.  The New Mexico farmer can assign a probability 
for rain based on his own experience and observations or, perhaps 
better, obtain a probability for rain from the National Weather 
Service or from private weather-forecasting concerns.  The experimenter 
with the polarizing beam splitter assigns probabilities to the two 
linear polarizations based on what he knows about the photon.  
Suppose, for example, that the beam splitter is oriented at 
$45^\circ$ to the vertical; the two output paths then correspond 
to linear polarizations at $45^\circ$ and $135^\circ$ to the vertical.  
If the experimenter knows that the photon is vertically polarized, 
then the rules of quantum mechanics dictate that he assign equal 
probabilities to the two output polarizations.  

These probabilities are {\it Bayesian probabilities}.$^{\hbox{\sevenrm 1--4}}$  
%Laplace1812,Jeffrey1961,Jaynes1989b,DeFinetti1990
Bayesian probabilities apply to a single realization of the alternatives,
as is evident in the examples.  They are subjective in that their 
values depend on what one knows.  Sometimes, as in the case of 
quantum mechanics, there are explicit rules for converting one's 
knowledge into a probability assignment; in other cases, the values 
of the probabilities represent little more than hunches.  

Reliable prediction requires probabilities that are close to 0 or 1.  
If probabilities for a single realization are not close to 0 or 1, 
it is still possible to make reliable predictions, {\it if\/} one has 
available a large ensemble of independent, identical realizations, 
all described by the same probabilities.   In such a large ensemble, 
the frequencies of occurrence of the various alternatives can be 
predicted reliably to be close to the probabilities.  Indeed, this is 
the reason that ensembles are used---to convert unpredictability for a 
single realization into reliable predictions for frequencies within a 
large ensemble of independent, identically distributed realizations.  

Because the scientific enterprise requires precise predictions, 
scientific experiments often use a large ensemble of independent, 
identically distributed realizations.  Sometimes, as in the case of 
the photon, such an ensemble is cheap, making it easy to perform 
precise experimental tests of probability assignments.  In other cases, 
as in clinical trials of new drugs, it is difficult to assemble an 
appropriate ensemble, the large cost being justified only by the need 
for a precise test.  Because of the importance of ensembles in 
scientific experiments, scientists are vulnerable to the notion 
that probabilities have meaning {\it only\/} as frequencies in a 
large ensemble.$^5$
%VonMises1921
Yet probabilities for a single realization are used routinely in 
rational decision making.  The example of the horse race shows this: 
the bettor determines his strategy for a single race from the 
probabilities that he assigns to that race.

The most misunderstood aspect of Bayesian probability theory is 
the relation between the probabilities for a single realization 
and frequencies within an ensemble of realizations.  Because this
relation is so important, yet so often misunderstood, it deserves 
a brief digression here.  An ensemble has a bigger 
space of alternatives than does a single realization.  
The probabilities on this larger space do not follow from the 
probabilities assigned to individual realizations within the ensemble.  
Assigning probabilities to the ensemble alternatives involves 
considerations beyond those needed to assign probabilities to 
individual realizations.$^{4,6}$
%DeFinetti1990,Jaynes1986  
In particular, if there are correlations among the members of the 
ensemble, the probabilities assigned to the ensemble should reflect 
these correlations.  When there are correlations, the probabilities 
for an individual realization generally do not predict frequencies 
within the ensemble.  A simple example is that of a biased coin, 
the bias known exactly except for its sign, which is unknown.  
The single-realization probabilities for heads and tails, which are 
clearly equal, do not predict the many-toss frequencies.  The lesson 
here is that {\it Bayesian probabilities for individual realizations 
do not necessarily predict frequencies within an ensemble of 
realizations}.

There are special cases, however, for which the probabilities for 
individual realizations {\it do\/} predict reliably the frequencies 
within an ensemble.  One of these special cases is the case of a large 
ensemble of independent, identically distributed realizations.  This 
successful prediction of frequencies, important though it is, should 
not lead to the erroneous conclusion that probabilities are equivalent 
to frequencies within a large ensemble.  Relations between probabilities 
and frequencies cannot be posited; they must be derived from probability 
theory itself.

The New Mexico farmer illustrates more generally how probabilities for
a single realization enter into rational decision making, providing 
a simple example of {\it decision theory}.$^{7,8}$
%Savage1972,Berger1985,
For simplicity, assume that to be effective the biological agent must be 
applied on a particular day during the development of the pest.  The 
Weather Service provides the farmer probabilities $\prain$, for rain 
that day, and $\pnorain=1-\prain$, for no rain.   If the farmer decides 
that it is not going to rain, he buys the biological agent and applies 
it to his chiles; if he decides that it is going to rain, he doesn't.  
Yet how is the farmer to decide?  Though the probabilities describe 
completely his uncertainty, he can't make a decision on the basis of the 
probabilities alone.  In addition to the probabilities, he must consider 
the costs of his actions.  Given the costs, he can use the probabilities 
to make a rational decision.  

Suppose that the pest, if left unchecked, reduces the value of the crop 
by $a$ dollars, and suppose that to buy and apply the biological agent 
costs $b$ dollars.  If the farmer doesn't buy the agent, then, rain or 
shine, his cost is the $a$ dollars for a damaged crop.  If the farmer 
buys and applies the agent, his cost is $a+b$ dollars if it rains and 
$b$ dollars if it doesn't; the average cost of buying and applying the 
agent is thus $(a+b)\prain+b\pnorain=a\prain+b$.  The farmer should 
take the action that has lower average cost.  Hence, he should buy 
and apply the agent if $a\prain+b<a$ or, equivalently, if $\pnorain>b/a$.  
The costs determine how confident the farmer should be that it will 
not rain before it is prudent for him to buy and apply the biological 
agent.  

Reliable prediction corresponds to probabilities near 0 or 1; other
probability distributions describe varying degrees of unpredictability.
How can the degree of unpredictability be quantified?  A natural 
measure comes from information theory.  The information missing toward 
specification of the actual alternative is given by the {\it Shannon 
information\/}$^{\hbox{\sevenrm 9--11}}$
%Shannon1949,Gallager1968,Cover1991
$$
H=
-\sum_j p_j\log_2 p_j
\;,
\eqno(1)$$
where $p_j$ is the probability for alternative~$j$ (the use of base-2 
logarithms means that information is measured in bits).  An alternative 
that has unit probability can be predicted definitely to occur; in 
this case there is no missing information, i.e., $H=0$.  At the other 
extreme, if the alternatives are equally likely, then the situation 
is maximally unpredictable; the missing information is maximal, with 
value $H=\log_2J\;$bits, where $J$ is the number of alternatives.  This 
is the case, for example, when the vertically polarized photon is 
incident on the beam splitter oriented at $45^\circ$; the equal 
probabilities for the two output paths correspond to 1 bit of 
missing information.  For probability distributions between the two 
extremes, the Shannon information takes on intermediate values.

One can acquire the missing information by observing which alternative 
is the actual one.  Wait for the finish of the horse race, and see which 
horse wins.  Observe the frisbee to see which path it takes.  Wait to see 
it if rains.  Wait for the photon to pass through the polarizing beam 
splitter, and determine which direction it goes.  This sort of observation
is, of course, not prediction.  It is worth stressing that the reason it
is not prediction is not the element of ``waiting'' in our examples; the 
reason is that the missing information is acquired by observing the 
alternatives themselves.  

In contrast to observing the alternatives, it is often the case that 
uncertainty about the actual alternative can be reduced or even 
eliminated entirely by gathering information about factors that 
influence the alternatives or determine the actual alternative.  For 
example, the frisbee golfer, by combining detailed observations of the 
wind velocity immediately upstream (initial conditions) with a model 
of the local terrain (boundary conditions), could integrate the coupled 
equations for the wind and the frisbee's motion, thus allowing him to 
predict with certainty the frisbee's path.  The New Mexico farmer might 
persuade the National Weather Service to gather sufficiently fine-grained 
meteorological data (initial and boundary conditions) so that it could 
integrate the hydrodynamic equations for the atmosphere, thereby 
predicting thunderstorms a day in advance.  In the spirit of these 
examples, it is tempting to posit in general an independent record of 
the missing information, outside the realization of interest, an 
``archive'' that records the actual alternative.  The archive stores 
the missing information in an encoded form, which must be decoded 
through, for example, integration of an appropriate partial differential 
equation.  Nonetheless, if one has access to the archive and can decode 
the information stored there, predictability can be restored.

It is useful to compare this description of unpredictability with that 
of a noiseless communication channel.$^{\hbox{\sevenrm 9--11}}$
%Shannon1949,Gallager1968,Cover1991
A transmitter prepares one of several alternative messages and sends 
it down a channel to a receiver, which reads the message.  How much 
information is communicated from transmitter to receiver?  The transmitter 
prepares the alternative messages with various probabilities.  The 
receiver is unable to predict which message it will receive and thus 
acquires, on average, an amount of information given by the Shannon 
information constructed from the message probabilities.  The transmitter 
retains a record of which message it sent, perhaps in an encoded form; 
by consulting the transmitter directly, one can eliminate the uncertainty 
about which message is sent down the channel.

The archive where a record of the actual alternative is stored is
like a transmitter: it is a source both of unpredictability and of 
information.  Indeed, the lesson is that {\it a source of unpredictability 
is the same as a source of information}.  By gaining access to the 
archive, one can acquire the missing information about which alternative 
is the actual one, thereby restoring predictability.  

In the examples of frisbee golf and the New Mexico farmer, the 
missing information is available in the initial conditions and boundary 
conditions that determine a unique solution to a set of differential 
equations.  In the horse-race example, the bettor can improve his 
prediction by gathering data about the previous performances of the 
horses and about specific conditions on the race day.  We don't know
if the remaining unpredictability can be eliminated by gathering yet
more information, because our understanding of the factors that enter
into determining the winner of a horse race is incomplete.  To be more
precise, we don't know whether there is a mathematical model that 
specifies what information needs to be collected and how that information 
is to be decoded so as to predict the winner with something approaching 
certainty.  Because we have no complete mathematical model of a horse 
race, we place its unpredictability, for the present, outside the scope 
of scientific inquiry.

The case of the photon is the most interesting.  Within quantum 
mechanics there is {\it no\/} archive that can be consulted to determine 
the photon's path through the beam splitter, {\it no\/} identifiable 
transmitter of the bit of information that specifies the photon's linear 
polarization.  The bit pops into existence out of nowhere.  Yet, unlike
the horse race, where there is no complete mathematical model, quantum
mechanics is thought to be a complete theory, which provides the framework 
for all fundamental physical law, a framework in which the probabilities 
are intrinsic.  Quantum-mechanical probabilities cannot be eliminated 
by gathering more information about the photon's state, for to say 
that the photon is vertically polarized is already a maximal quantum 
description of its state.  

We are in a position now to characterize how fundamental a source of 
unpredictability is.  It seems sensible to say that the more difficult 
it is to consult the archive and acquire the missing information necessary
for predictability, the more fundamental is the unpredictability.  In 
this regard quantum unpredictability is in a class by itself: there is 
{\it no\/} archive that stores an independent record of the information 
that is acquired in a quantum measurement.  Quantum unpredictability is
a consequence of information without a source; it cannot be eliminated 
by consulting an archive because there is no archive.  Unpredictability 
without an information source is so fundamental that we reserve for it 
the appellation {\it absolute unpredictability}.

In the case of the New Mexico thunderstorm, the missing information is 
available as initial data, but it is is very difficult to obtain because 
of the phenomenon of {\it classical chaos}, which means that the 
coarse-grained past does not predict the coarse-grained future.  To 
predict a phenomenon on the coarse-grained scale of a thunderstorm 
requires initial data on a much finer scale---indeed, a scale that is 
exponentially finer in the time over which one desires a reliable
prediction.  

Although the data required to predict the coarse-grained dynamics 
of a chaotic system are very difficult to obtain, that they are 
obtainable in principle cautions that care should be exercised in 
characterizing the difficulty of the task.  In the spirit of decision 
theory, what ought to be done is to compare costs: the cost of consulting 
the archive, obtaining the missing information, and thereby eliminating 
the unpredictability should be compared with the cost of not having 
the missing information and thereby having to deal with the resulting 
unpredictability.  If the cost of obtaining the missing information 
exceeds the benefit of having it, we can point to a fundamental reason 
for ``allowing'' the unpredictability.  Indeed, a direct comparison of costs 
is the only way we know of to quantify how fundamental a source of 
unpredictability is.  We refer to unpredictability for which the cost 
of obtaining the missing information far exceeds the benefit of having 
it as {\it strong unpredictability}.  Having introduced the notions of 
absolute unpredictability and strong unpredictability, we can continue 
to use---and encourage others to use---the phrase ``fundamental 
unpredictability'' in any other fashion desired.

We develop these ideas in the next section, with a discussion of three
sources of unpredictability in physics: the absolute unpredictability 
of quantum mechanics, the unpredictability of the coarse-grained future
of a classically chaotic Hamiltonian system, and unpredictability that 
arises when a physical system is coupled to a perturbing environment.  
Our main interest is how these three kinds of unpredictability are related 
to chaos in classical and quantum dynamics.  The first two sources of 
unpredictability have already been discussed; their incompatibility 
lies at the heart of the difficulty in formulating a description of 
chaos in quantum dynamics.  A further difficulty is that the 
unpredictability of the 
coarse-grained future of a chaotic system does not lend itself to a 
meaningful comparison of the costs and benefits of obtaining the missing 
information.  The third kind of unpredictability, due to environmental 
perturbation, can be used to put chaos in classical and quantum dynamics 
on the same footing, as we show in the last section of the article.  
In particular, we describe a new characterization of classical chaos 
in terms of sensitivity to environmental perturbation, a characterization 
in which costs and benefits can be compared directly, with classical 
chaos emerging as a strong source of unpredictability.  We indicate 
briefly how this same way of characterizing chaos can be applied to 
quantum dynamics.
   
\section{II. \ SOURCES OF UNPREDICTABILITY IN PHYSICS}

In this section we focus on scientific unpredictability, specifically,
unpredictability in physics.  One doesn't have to look far to find 
such unpredictability; we need only look for any place where physicists 
employ a probabilistic description.

The obvious place to look is quantum mechanics, which in our present
understanding provides the framework in which fundamental physical
laws are formulated.  In a quantum-mechanical description, even if one 
has maximal information about a physical system, i.e., knows its quantum 
state exactly, one nonetheless cannot predict the results of most 
measurements.  We can extend the photon example to provide a vivid 
illustration of this.  Suppose that the photon, initially polarized 
along the vertical axis, is incident on a series of polarizing beam 
splitters whose orientations alternate between $45^\circ$ and vertical.  
At each beam splitter there is a bit of missing information about which 
direction the photon goes.  It appears that the photon is an 
inexhaustible source of information, yet within the conventional 
formulation of quantum mechanics, there is no source for this 
information, no archive that can be consulted to predict which path 
the photon will take through the series of beam splitters.  This is
what we have designated an absolute source of unpredictability.

One might expect the entire discussion in this Workshop to be focused 
on the absolute unpredictability of quantum mechanics.  That it isn't
requires explanation, and we can suggest two reasons.  The first is 
that an alternative theory in which there is an identifiable source for 
the missing information has serious drawbacks, a fact made evident by 
30 years of work on Bell inequalities.$^{12}$
%Peres1993
The desire for a source for the missing information---a quantum 
archive---is strong.  From the early days of quantum mechanics, 
many physicists have found it unreasonable to have intrinsic 
unpredictability---unpredictability without a source of 
information---and they have posited the existence of ``hidden 
variables'' that constitute a quantum archive.$^{12,13}$
%Peres1993,Bell1987
The hidden variables, though they are presently and perhaps permanently 
inaccessible, would provide enough information, could they be consulted, 
to eliminate quantum unpredictability.  

The catch is the following: the Bell inequalities show that if a 
hidden-variable theory is to agree with the statistical predictions of 
quantum mechanics---and, as experiments show, if it is to agree with 
observation---the hidden variables must be nonlocal.  The archive 
cannot have independent subarchives for different subsystems (for 
example, a subarchive within the apparatus that observes each subsystem),
but rather must be one enormous record that commands and correlates 
the behavior of everything in the Universe.%
\footnote{$^{\hbox{\sevenrm [1]}}$}
{This conclusion about the nonlocality of a hidden-variable archive 
is true even if the archive is forever inaccessible.  Indeed, the 
notion of a nonlocal archive is perhaps easier to swallow if it 
is inaccessible.}\  
The necessity for hidden-variable theories to be nonlocal makes them 
considerably less attractive---depending on one's taste, even less 
attractive than the absolute unpredictability of quantum mechanics.  
Yet if it turns out that the fundamental constituents of matter exist 
in more dimensions than the four of our familiar spacetime, as in 
string theory, then locality within four-dimensional spacetime might 
lose much of the force we presently attach to it.  

The second reason, mentioned in Hartle's introductory lecture at
this Workshop, is this: the present Universe is enormously complex, 
its particularities describable only by a great deal of information; 
if the fundamental physical laws and the initial conditions are simple, 
where does all that information come from?  In a hidden-variable theory, 
the complexity of the present Universe is a revelation of the details 
of the hidden variables; because the hidden variables can be thought 
of as part of the initial conditions, the initial conditions necessarily 
become complex.  It is somehow more appealing to imagine that the laws 
and initial conditions are simple and that there is no archive in 
which is written an independent record of the complexity of the 
present Universe; quantum mechanics obliges by making the complexity 
almost wholly a consequence of the unpredictability of quantum rolls 
of the dice.

What about unpredictability in classical physics?  Nonlinear classical
systems---here restricted to Hamiltonian systems---can display a kind
of unpredictability that comes from classical chaos.$^{14}$
%Lichtenberg1983
Classical chaos is usually characterized in terms of the unpredictability
of phase-space trajectories.  Consider the points along a phase-space 
trajectory at a discrete sequence of uniformly spaced times.  The points 
are never given to infinite precision; any finite precision corresponds 
to a gridding of phase space into coarse-grained cells.  The sequence 
of finite-precision points, coarse grained both on phase space and in 
time, is what is meant by a coarse-grained trajectory.  For a classically 
chaotic system, coarse-grained initial data do not predict a unique 
coarse-grained trajectory; more precisely, to predict a unique 
coarse-grained trajectory requires initial data that become exponentially 
finer in the time over which the prediction is desired.

It is instructive to review the mathematical formulation of classical 
chaos in which the initial data appear explicitly as a source of 
unpredictability and information.  Consider a classical system whose 
motion is restricted to a compact phase-space region, represented as
a square in Figure~1.  Grid this phase space into coarse-grained cells 
of uniform volume ${\cal V}$.  The coarse-grained initial data (at time 
$t=0$) are that the initial phase-space point lies somewhere within a 
particular coarse-grained cell.  This corresponds to a phase-space 
density that is uniform on the initial cell.  Under a chaotic Hamiltonian 
evolution, the phase-space density spreads across phase space, creating 
a pattern of uniform density (see Figure~1), which occupies the same 
volume as the initial cell and which develops structure on finer and 
finer scales as the evolution proceeds.

At each of the discrete times $t$, the evolved pattern can be partitioned 
into all its separate intersections with the initial grid.  Each piece of 
the partition corresponds to at least one coarse-grained trajectory that 
issues from the initial cell and terminates in that piece at time $t$.  
It turns out that we introduce no error into the present discussion by 
pretending that each piece of the partition corresponds to a {\it unique\/} 
coarse-grained trajectory.  Label the various pieces by an index $j$, and 
let ${\cal V}_j$ be the phase-space volume of the $j$th piece.  The 
probability for the corresponding coarse-grained trajectory is 
$q_j={\cal V}_j/{\cal V}$, the fraction of the original phase-space 
volume occupied by the $j$th piece.  The information needed to specify 
a particular coarse-grained trajectory out to time $t$ is given by the 
Shannon information constructed from the probabilities $q_j$.  For a 
chaotic evolution, in the limit of large $t$, this information grows 
linearly in time:
$$
-\sum_jq_j\log_2q_j\sim 
Kt
\;.
\eqno(2)$$
The linear rate of information increase, $K$, called the 
{\it Kolmogorov-Sinai\/} (KS) or {\it metric entropy},$^{15}$
%Alekseev1981
quantifies the degree of classical chaos.%
\footnote{$^{\hbox{\sevenrm [2]}}$}
{The definition of KS entropy given here is not quite right, because 
$K$ can depend on the choice of initial cell.  The quantity that grows 
asymptotically as $Kt$ is really the average of the information on 
the left side of Eq.~(2) over all initial cells.  We ignore this 
distinction here, thereby assuming implicitly that the chaotic system 
has roughly constant Lyapunov exponents over the accessible region 
of phase space.}

The information~(2) is missing information about which is the actual
coarse-grained trajectory.  The missing information can be obtained from 
the actual initial condition.  The correspondence can be made explicit 
in the following way.  The $j$th piece of the partition of the evolved 
pattern corresponds to a region of volume ${\cal V}_j$ within the initial 
cell; this region, which has probability $q_j$, is the region of initial 
conditions that lead to the coarse-grained trajectory that terminates 
in the $j$th piece of the partition.  Thus at each time $t$ the initial 
cell is partitioned into initial-condition regions, each of which gives 
rise to a particular coarse-grained trajectory out to time $t$.

Imagine gridding the initial cell into very fine cells of uniform volume 
$\Delta v$, cells so fine that they are much finer than the 
initial-condition regions for all times of interest.  The information 
needed to specify a particular fine-grained cell within the initial 
coarse-grained cell---this is the {\it entropy\/} of the initial 
phase-space density---is $\log_2({\cal V}/\Delta v)$.  This information, 
when written as
$$
\log_2{{\cal V}\over\Delta v}=
-\sum_jq_j\log_2q_j+
\sum_jq_j\log_2{{\cal V}_j\over\Delta v}
\;,
\eqno(3)$$
illustrates how the initial data act as an archive for the coarse-grained
trajectory.  The first term on the right is the information needed to 
specify the initial-condition region for a particular coarse-grained 
trajectory.  The second term is the further information needed to specify 
a fine-grained cell within an initial-condition region.  The total 
information needed to specify a fine-grained cell is the sum of these 
two terms.  As a chaotic evolution proceeds, more and more of the 
information needed to specify a fine-grained initial cell is required to 
predict a particular coarse-grained trajectory.  

A crude, but instructive picture of what is happening is that the number 
of pieces in the partition of the evolved pattern grows as $2^{Kt}$, 
each piece having roughly the same phase-space volume and, hence, the 
same probability $q_j=2^{-Kt}$.  As the evolution proceeds, the 
corresponding partition of the initial cell becomes exponentially finer, 
consisting of roughly $2^{Kt}$ initial-condition regions, and the 
information needed to specify a particular region grows linearly in time.  
A coarse-grained trajectory can be regarded as a progressive unveiling 
of finer and finer details of the actual initial data within the 
initial coarse-grained cell. 

How fundamental is the chaotic unpredictability of coarse-grained 
trajectories?  It's not absolute unpredictability, as in quantum 
mechanics, because it is easy to identify the source of information 
that must be consulted to eliminate the unpredictability.  The 
unpredictability of a coarse-grained trajectory is due wholly to a 
lack of knowledge of the initial conditions.  The source of information 
is thus the initial conditions, which when decoded through the equations 
of motion, yield a unique prediction for the coarse-grained trajectory.  
Of course, if the required initial data are so fine that they are at 
the quantum level on phase space, then the unpredictability of the 
coarse-grained trajectory becomes sensitive to the absolute 
unpredictability of quantum mechanics.  Classical chaos then serves 
as an amplifier of quantum unpredictability to a classical level.$^{16}$  
%Fox1990

Suppose we wish to assess how fundamental is the unpredictability of 
a chaotic coarse-grained trajectory.  Since the initial data are a source
for the missing information needed to predict a coarse-grained trajectory,
we ought to compare the cost of obtaining the necessary initial data 
with the cost of the unpredictability that comes from not having the
required data.  Here a problem arises: it is difficult to formulate this 
comparison in a way that is intrinsic to the system under consideration, 
because the costs generally depend on factors external to the system.  
Take the New Mexico farmer as an example.  The cost of acquiring the 
required initial data depends on the level of technology used in gathering 
the meteorological data.  Worse, the cost of unpredictability is highly 
dependent on who assesses the cost.  It may be important to the farmer 
to know whether it will rain, but the one of us who lives in Albuquerque 
generally doesn't care much whether there is a thunderstorm on a 
particular summer day.  When he does care, a ten-minute warning, 
easily obtained by looking out the window, is generally sufficient.

Classical physics has none of the absolute unpredictability of quantum 
mechanics.  Does quantum mechanics have any of the sensitivity to initial 
conditions that is displayed by classically chaotic systems?  There is 
no sensitivity to initial conditions in the evolution of the quantum 
state vector: the unitarity of quantum evolution implies that it 
preserves inner products, so the ``distance'' between state vectors 
remains unchanged during quantum evolution.  Suppose one looks for 
sensitivity to initial conditions in the ``coarse-grained trajectory'' 
of some observable like position or momentum.  Such a coarse-grained 
quantum trajectory is constructed by periodically making coarse-grained 
measurements of the observable.  The problem is that the measurements 
generally introduce the absolute unpredictability of quantum mechanics, 
making the coarse-grained trajectory unpredictable for reasons that are 
essentially independent of the quantum dynamics.  One ends up studying 
not any sort of sensitivity to initial conditions in quantum dynamics, 
but rather the absolute unpredictability of quantum mechanics.  

The incompatibility of the absolute unpredictability of quantum mechanics 
with the classical unpredictability due to sensitivity to initial 
conditions is the chief difficulty in formulating a description of 
quantum chaos.  What is needed is a description of chaos that, avoiding 
the absolute unpredictability of quantum mechanics and the classical 
sensitivity to initial conditions, is formulated in terms of a form 
of unpredictability that is common to classical and quantum physics.  
Notice, in particular, that instead of trying first to formulate a 
description of quantum chaos, the primary task is to reformulate the 
description of classical chaos, albeit in a way that is equivalent to 
the standard characterization in terms of the unpredictability of 
coarse-grained trajectories.  We have suggested and investigated such 
a new way to characterize chaos, which we introduce here by considering 
yet a third source of unpredictability, the unpredictability of an 
open system, i.e., a system that is coupled to a perturbing 
environment.$^{\hbox{\sevenrm 17--22}}$
%Caves1993b,Schack1992a,Schack1993e,Schack1994b,Schack1996a,Schack1996b

In investigating this third source of unpredictability, an essential 
tool is the entropy of a physical system.  We introduce the notion 
of entropy, in both classical and quantum physics, as the missing 
information about the system's fine-grained state.$^{23,24}$
%Jaynes1957a,Jaynes1957b
For a classical system, suppose that phase space is gridded into very 
fine-grained cells of uniform volume $\Delta v$, labeled by an index 
$j$.  If one doesn't know which cell the system occupies, one assigns 
probabilities $p_j$ to the various cells; equivalently, in the limit 
of infinitesimal cells, one can use a phase-space density 
$\rho(X_j)=p_j/\Delta v$.  The {\it classical entropy\/} (measured 
in bits), 
$$
H=
-\sum_j p_j\log_2 p_j=
-\int dX\,\rho(X)\log_2\bigl(\rho(X)\Delta v\bigr)
\;,
\eqno(4)$$
is the missing information about which fine-grained cell the system 
occupies.  For example, throughout this article we use as initial
data a phase-space density that is uniform on a coarse-grained cell
of volume ${\cal V}$; the corresponding entropy is 
$\log_2({\cal V}/\Delta v)$.  In quantum mechanics the fine-grained 
alternatives are normalized state vectors in Hilbert space.  From a 
set of probabilities for various state vectors, one can construct a 
density operator
$$
\hat\rho=\sum_j\lambda_j|\psi_j\rangle\langle\psi_j|
\;,
\eqno(5)$$
where the state vectors $|\psi_j\rangle$ are the eigenvectors of 
$\hat\rho$, with eigenvalues $\lambda_j$.  The normalization of the 
density operator, ${\rm tr}(\hat\rho)=1$, implies that the eigenvalues 
make up a normalized probability distribution.  The {\it von Neumann 
entropy\/} of $\hat\rho$ (measured in bits), 
$$
H=
-{\rm tr}(\hat\rho\log_2\hat\rho)=
-\sum_j\lambda_j\log_2\lambda_j
\;,
\eqno(6)$$
can be thought of as the missing information about which eigenvector 
the system is in.  

Entropy quantifies the degree of unpredictability about the system's
fine-grained state.  What makes it such an important quantity is that
there is a readily identifiable cost, intrinsic to the system, for the
inability to predict the system's fine-grained state.  Suppose that 
the system exists in the presence of a heat reservoir at temperature 
$T$, so that all exchanges of energy with the system that are not in 
the form of useful work must ultimately be exchanged with the reservoir 
as heat.  Then each bit of entropy reduces the useful work that can 
be extracted from the system by $k_BT\ln2$, where $k_B$ is the Boltzmann 
constant.  (The factor $k_B\ln2$ is a change of units; it translates 
entropy from bits to conventional thermodynamic units.)  The cost of 
missing information is a reduction in the useful work that can be 
extracted from the system.

Entropy remains unchanged under Hamiltonian dynamical evolution, both 
classically and quantum mechanically.  Classically this follows from
the preservation of phase-space volume under Hamiltonian evolution;
quantum mechanically it follows from the unitarity of Hamiltonian
evolution, which preserves the eigenvalues of the density operator.
Suppose, however, that the system is coupled to a perturbing environment.
The interaction disturbs the system's evolution; averaging over the 
disturbance generally causes the system's entropy to increase.  This 
is a standard mechanism for entropy increase, the increase quantifying 
the decreasing ability to predict the fine-grained state of the system.  
In this case it is obvious that the source of unpredictability---the 
source of information---is the perturbing environment.  By observing 
the environment, one can determine aspects of the perturbation, thus 
reducing the entropy increase of the system and rendering the 
fine-grained state of the system more predictable.  

The rub is that the information acquired by observing the environment 
has a thermodynamic cost, too, a cost paid when the information is 
erased.  For erasure into a heat reservoir at temperature $T$, this 
{\it Landauer erasure cost\/}$^{25,26}$
%Landauer1961,Landauer1988
is $k_BT\ln2$ per bit, exactly the same as the cost of missing 
information.  The Landauer erasure cost exorcises Maxwell 
demons.$^{\hbox{\sevenrm 27--30}}$
%Bennett1982,Zurek1989a,Zurek1989b,Caves1989
A demon observes a system directly, thereby decreasing the system's 
entropy---according to the demon, the system's fine-grained state 
becomes more predictable---and increasing the amount of work that 
the demon can extract from the system.  The demon can't win, however, 
because the entropy reduction, averaged over the possible results 
of the observation, is equal to the amount of information acquired 
from the observation; hence the erasure cost of the acquired 
information cancels the increase in available work.  Turned on its 
head, this line of argument shows that if the Second Law of 
Thermodynamics is to be maintained against the demon's attack, 
acquired information {\it must\/} have a thermodynamic cost of 
$k_BT\ln2$ per bit, as was first noted by Szilard$^{31}$;
%Szilard1929
Landauer$^{25}$
%Landauer1961
realized that the cost is paid when the information is erased.

A demon observes the system directly.  Here we contemplate something
different: making inferences about the system by observing the environment 
that interacts with it.  This difference is crucial for two reasons.  
First, when observing the environment, there is no necessary balance 
between the entropy reduction and the amount of acquired information; 
this permits a nontrivial comparison between the cost of acquiring the 
information from the environment and the cost of not having it.  Second, 
by observing only the environment, we are considering a kind of 
unpredictability that can be formulated in the same way in both classical 
and quantum physics; this allows a meaningful comparison of classical 
and quantum dynamics.  Both these reasons deserve further discussion.

To discuss the first reason, it is useful to introduce the notation 
that we use in comparing costs.  Averaging over the perturbing 
environment causes the system's entropy to increase by an amount 
$\Delta H_0$.  By observing the environment, one can make the system's 
entropy increase smaller.  We let $\Delta I$ be the amount of information 
acquired from the observation, and we let $\Delta H\le\Delta H_0$ be 
the corresponding entropy increase of the system, averaged over the 
possible results of the observation.  The reduction in the system's 
entropy as a consequence of the observation is $\Delta H_0-\Delta H$.  
The acquired information, which has a thermodynamic cost of 
$\Delta I\,k_BT\ln2$, buys an increase in available work of 
$(\Delta H_0-\Delta H)k_BT\ln2$.  Because entropy and acquired information 
weigh the same in the balance of thermodynamic cost, we can compare 
directly the cost of acquiring the information with the benefit of 
having it just by comparing the amount of acquired information, 
$\Delta I$, with the entropy reduction it purchases, 
$\Delta H_0-\Delta H$.  

If one doesn't observe the environment, one acquires no information,
i.e., $\Delta I=0$, and the entropy reduction is zero, i.e., 
$\Delta H=\Delta H_0$.  A very coarse observation of the environment 
gathers very little information and yields very little entropy reduction.  
The entropy reduction can be made progressively larger by making 
progressively more detailed observations, which gather more and more 
information about the environmental perturbation.  The entropy reduction 
cannot exceed the information acquired, i.e., 
$$
\Delta I\ge 
\Delta H_0-\Delta H
\eqno(7)
$$
\hbox{}---this is another expression of the Second Law---but generally 
the entropy reduction is smaller, and can be much smaller, than the 
information acquired from the observation.  

How can such an imbalance occur?  Classically, the reduction in 
entropy that comes from observing the environment can be pictured 
roughly in the following way.  Averaging over the perturbing environment 
yields a phase-space density---this we call the {\it average 
density}---that occupies a phase-space volume bigger than the initial 
volume by a factor of $2^{\Delta H_0}$.  The observation determines a 
phase-space density, within the average density, that occupies a volume 
smaller than the average density by a factor of $2^{\Delta H_0-\Delta H}$.  
If the results of the observation corresponded to a set of (equally likely) 
{\it nonoverlapping\/} densities that fit within the average density,
then, there being about $2^{\Delta H_0-\Delta H}$ of these nonoverlapping 
densities, the information acquired from the observation would be roughly 
$\log_2(2^{\Delta H_0-\Delta H})=\Delta H_0-\Delta H$.  Generally,
however, the results of the observation correspond to {\it overlapping\/} 
densities within the average density, of which there can be many more 
than the number of nonoverlapping densities.  Consequently, the 
information $\Delta I$ can be much larger than the entropy reduction.  
The discussion in the next section indicates how a proliferation of 
overlapping perturbed densities arises as a consequence of chaotic 
classical dynamics, the result being an exponential imbalance between 
information and entropy reduction.  

The same explanation for a potential imbalance between acquired information
and entropy reduction works quantum mechanically, with the average 
phase-space density replaced by an average density operator and with 
the nonoverlapping and overlapping densities replaced by orthogonal 
and nonorthogonal density operators.  The potential for an imbalance in 
quantum mechanics arises because an observation of the environment 
generally determines one of a set of {\it nonorthogonal\/} density 
operators, of which many more can contribute to the average density 
operator than can orthogonal density operators.

For open systems strong unpredictability can now be seen to mean
that the cost of acquiring information from the environment and thus 
making the fine-grained state more predictable is much greater than 
the benefit of having that predictability, i.e.,
$$
\Delta I\gg 
\Delta H_0-\Delta H 
\;.
\eqno(8)
$$
We say that a system is {\it hypersensitive to perturbation\/} if,
for the optimal way of observing the environment, it displays this 
strong unpredictability for all values of $\Delta H$.  We have used 
hypersensitivity to perturbation to characterize classical and quantum 
chaos.  A rigorous analysis$^{21}$ in symbolic dynamics$^{15}$
%Schack1996a
shows that classically chaotic systems display an {\it exponential 
hypersensitivity to perturbation}, for which 
$$
\Delta I\sim 
2^{Kt}(\Delta H_0-\Delta H)
\;.
\eqno(9)
$$
The acquired information becomes exponentially larger than the 
entropy reduction, with the exponential rate of increase given by 
the KS entropy of the chaotic dynamics.  In the next section we 
present a heuristic version of the symbolic dynamics analysis.

What allows one to compare costs directly, in a way that is intrinsic 
to the system, is the connection of missing and acquired information 
to thermodynamics and statistical physics.  Indeed, this connection 
provides a statistical-physics motivation for our approach.  Why does 
one average over the environment and allow the entropy of a system to 
increase?  Usually one gives an excuse: the environment is said to
be so complicated that averaging over it is the only practical way to 
proceed.  We reject such apologetics in favor of a direct comparison 
of costs.  When a system is hypersensitive to perturbation, so that 
the acquired information far exceeds the entropy reduction, it is 
thermodynamically highly unfavorable to try to reduce the system 
entropy by gathering information from the environment.  The 
thermodynamically advantageous course is to average over the perturbing 
environment, thus allowing the system entropy to increase.  Thus 
{\it the strong unpredictability of classically chaotic open systems 
provides a justification for the entropy increase of the Second Law
of Thermodynamics}.

Return now to the second reason for considering observations of the 
environment.  Classically we are dealing with the ability to predict 
a phase-space density when a system is disturbed by a perturbing 
environment; quantum mechanically we are dealing with the ability to 
predict a state vector when a system is so disturbed.  Thus we deal 
with a kind of unpredictability that is common to classical and quantum
physics. 

The key to placing classical and quantum unpredictability on the same 
footing is to put aside phase-space trajectories, dealing instead with 
phase-space densities classically and with state vectors quantum 
mechanically.  When the system is unperturbed, the evolution of 
classical phase-space densities is governed by the Liouville equation, 
and the evolution of quantum state vectors by the Schr\"odinger equation.  
Neither the Liouville equation nor the Schr\"odinger equation displays 
sensitivity to initial conditions: the overlap of phase-space densities 
is preserved by the canonical transformations of Liouville evolution, 
and the overlap of state vectors is preserved by the unitary 
transformations of Schr\"odinger evolution.  Moreover, there is none 
of the absolute unpredictability of quantum mechanics, because we are 
considering deterministic Schr\"odinger evolution rather than the 
unpredictable results of measurements on the system.  To this predictable 
unperturbed evolution, we add unpredictability by including a perturbing 
environment, and we ask how the perturbation degrades the ability to 
predict phase-space densities classically and to predict state vectors 
quantum mechanically.  The source of unpredictability in both cases 
is the disturbance introduced by interaction with the environment.

Before going on to our heuristic argument for classical hypersensitivity,
it is important to mention that we model the perturbing environment by 
adding a stochastic term to the system Hamiltonian.  Such a stochastic 
perturbation can be realized as a particular kind of coupling to an 
environment: one couples the system to conserved quantities of an 
environment; different values of the conserved quantities specify the 
various realizations of the stochastic perturbation.  Such a stochastic 
Hamiltonian is by no means the most general kind of coupling to an 
environment.  What is missed by the stochastic model are classical 
correlation with the environment and quantum entanglement with the 
environment.

\section{III. \ UNPREDICTABILITY AND CHAOS}

We turn now to a discussion of hypersensitivity to perturbation in 
classically chaotic systems.  Our objective is not to give a rigorous 
analysis, but rather to capture the flavor of why classically chaotic 
systems display exponential hypersensitivity to perturbation.  For 
systems that have a symbolic dynamics,$^{15}$ a rigorous analysis 
has been given, and the reader intent on rigor or just interested 
in a more thorough formulation of the problem is referred to that 
analysis.$^{21}$ 
%Schack1996a

Consider a classical Hamiltonian system, which is globally chaotic on
a compact region of phase space, the degree of chaos characterized by 
the KS entropy $K$.  Recall the picture of the system dynamics that was
introduced in Section~II (see Figure~1).  The initial data, that the 
system occupies a coarse-grained cell of volume ${\cal V}$, correspond 
to an initial phase-space density that has the uniform value 
${\cal V}^{-1}$ over the initial cell.  Under the chaotic dynamics the 
initial coarse-grained cell is stretched and folded to form an 
increasingly intricate pattern on phase space.  The evolving pattern 
has the same volume ${\cal V}$ as the initial coarse-grained cell, and 
the phase-space density is uniform over the evolving pattern, with value 
${\cal V}^{-1}$.

The dynamics of the pattern can be described crudely as an exponential 
expansion in half the phase-space dimensions and an exponential 
contraction in the other half of the phase-space dimensions.  The 
exponential rate of expansion or contraction in a particular phase-space 
dimension is given by a typical Lyapunov exponent $\lambda=K/D$, where 
$2D$ is the dimension of phase space.  The expansion in the expanding 
dimensions means that the phase-space pattern spreads over roughly 
$(2^{\lambda t})^D=2^{Kt}$ coarse-grained cells at time $t$.  
The width of the pattern in a contracting dimension is roughly 
$2^{-\lambda t}{\cal V}^{1/2D}$.

We now imagine perturbing this evolution stochastically.  The perturbation
is modeled as a diffusion on phase space, characterized by a diffusion 
constant ${\cal D}$.  Such a perturbation is described by adding a 
stochastic term to the system Hamiltonian: during each small time
interval the system evolves according to its own Hamiltonian plus an 
additional Hamiltonian selected randomly from a continuous set of 
possible perturbing Hamiltonians.  For a particular temporal sequence of 
perturbing Hamiltonians---we call such a realization of the perturbation 
a {\it perturbation history\/}---the phase-space pattern is disturbed in 
a particular way (see Figure~1).  The resulting perturbed pattern is 
different from the unperturbed pattern, but it occupies the same volume 
${\cal V}$ as the unperturbed pattern, and the perturbed phase-space 
density has the uniform value ${\cal V}^{-1}$ on the perturbed pattern.  

Averaging over the possible perturbed patterns---i.e., averaging over 
all the perturbation his\-to\-ries---yields an {\it average\/} 
phase-space density, as shown in Figure~1.  The diffusion ``smears out'' 
the unperturbed pattern into an average density that occupies a larger 
volume than the unperturbed pattern.  

Formally the evolution of the phase-space density is described by a 
Liouville equation that has a stochastic contribution to the Hamiltonian.  
Each perturbation history corresponds to a particular realization 
of the stochastic term in this Liouville equation and yields a 
particular perturbed pattern.  The equation that governs the evolution 
of the average density is obtained by averaging the stochastic Liouville
equation over all perturbation histories.  The resulting evolution 
equation, a Fokker-Planck equation on phase space, has a systematic 
term that describes the unperturbed Hamiltonian evolution and a 
diffusion term that describes the perturbation.  The perturbation 
is characterized by its strength and by how it is correlated across 
phase space.  Both of these aspects of the perturbation are important 
for our discussion. 

An important time emerges from the interplay between the unperturbed 
dynamics and the diffusion.  During a typical Lyapunov time 
$\lambda^{-1}$, the diffusion smears out the average density by an 
amount $\sqrt{{\cal D}/\lambda}$ in each phase-space dimension.  In 
the expanding dimensions this smearing is overwhelmed by the expansion,
but in the contracting dimensions it becomes important once the width 
of the unperturbed phase-space pattern becomes comparable to the amount 
of diffusion in a Lyapunov time.  After this time the diffusion balances 
the exponential contraction of the dynamics, with the result that the 
average density ceases to contract in the contracting dimensions.  We 
say that the perturbation becomes {\it effective\/} at a time given 
roughly by
$$
\sqrt{{\cal D}/\lambda}\sim
2^{-\lambda\teff}\,{\cal V}^{1/2D}
\qquad\Longleftrightarrow\qquad
\teff\sim
{1\over2\lambda}
\log_2\!\left({\lambda{\cal V}^{1/D}\over{\cal D}}\right)
\;.
\eqno(10)
$$
No matter how weak the perturbation, the exponential contraction of the 
chaotic dynamics eventually renders the perturbation effective, typically
within several Lyapunov times.

After the perturbation becomes effective, the average phase-space 
density continues to expand exponentially in the expanding dimensions, 
but this expansion is no longer balanced by contraction in the contracting 
dimensions.  Thus the average density occupies an exponentially increasing
phase-space volume ${\cal V}_0\sim2^{K(t-\teff)}{\cal V}$---i.e., a factor
of $2^{K(t-\teff)}$ larger than the phase-space volume occupied by the 
unperturbed pattern---and the entropy of the average density increases
as
$$
\Delta H_0\sim 
K(t-\teff)
\quad\hbox{for $t\gsim\teff$.}
\eqno(11)
$$
Once the perturbation becomes effective, the entropy increase 
$\Delta H_0$ of the average density is determined by the KS entropy of 
the system dynamics, not by some property of the perturbation.
We assume for the remainder of the discussion that $t\gsim\teff$.

The unpredictability quantified by the entropy increase~(11) has an
obvious source in the stochastic perturbation.  The perturbation 
histories constitute an archive that can be consulted to reduce 
the entropy increase.  Acquiring information about the perturbation 
means finding out something about which perturbation history is the 
actual one.  Our task is to estimate how much information $\Delta I$ 
about the perturbation is required to reduce the entropy increase to 
$\Delta H\le\Delta H_0$.  Determining the actual perturbation history 
specifies a particular perturbed pattern, which, since it occupies the 
same phase-space volume as the unperturbed pattern, has the same entropy, 
so that the entropy increase $\Delta H$ is kept to zero.  For a 
diffusive perturbation, however, determining the actual perturbation 
history means singling out one history from an infinite number of 
histories and thus requires an infinite amount of information.  We are 
more interested here in acquiring {\it partial\/} information about the 
perturbation, which means determining that the actual perturbation 
history lies in some class of perturbation histories.  Suppose the 
corresponding class of perturbed patterns, when averaged together, 
produces a {\it partial\/} density that occupies a phase-space volume 
$2^{\Delta H}{\cal V}=2^{-(\Delta H_0-\Delta H)}{\cal V}_0$, so that
the corresponding entropy increase is $\Delta H$.  We must estimate how 
much information is required to determine that the actual perturbation 
history lies in such a class.  

To make such an estimate, we need to say something about how the 
diffusive perturbation is correlated across phase space.  For simplicity, 
let us assume that the perturbation is well correlated across a 
coarse-grained cell, but is essentially uncorrelated across scales 
larger than a coarse-grained cell.  At any particular time, the main 
effect of the perturbation is the diffusion during the last Lyapunov 
time, because the effects of the perturbation more than a few Lyapunov 
times in the past are suppressed by the exponential contraction.  To 
keep the entropy increase to $\Delta H$, one must acquire enough 
information about the perturbation histories so that the corresponding 
partial density occupies a phase-space volume that is a factor of 
$2^{\Delta H_0-\Delta H}$ smaller than the volume occupied by the 
average density.  

Now we come to the key point.  Within each coarse-grained cell occupied 
by the unperturbed pattern at time $t$, there are $2^{\Delta H_0-\Delta H}$ 
possible ``slots'' for the partial density.  Since the perturbation 
is essentially independent from one coarse-grained cell to the next, 
one must acquire enough information, in each coarse-grained cell, to 
determine in which of these slots the perturbed pattern actually lies.  
This means acquiring $\Delta H_0-\Delta H$ bits of information about 
the stochastic perturbation {\it for each coarse-grained cell occupied 
by the unperturbed pattern at time $t$}.  Since the unperturbed pattern 
spreads over about $2^{Kt}$ coarse-grained cells at time $t$, the total 
amount of information that must be acquired to keep the entropy increase 
to $\Delta H$ is roughly
$$
\Delta I\sim 
2^{Kt}(\Delta H_0-\Delta H)
\quad\hbox{for $t\gsim\teff$.}
\eqno(12)
$$

Equation~(12) is the main result of our heuristic argument.  Its 
content is the following: once the perturbation becomes effective, 
a classically chaotic system displays an exponential hypersensitivity 
to perturbation.  A simple example of the argument leading to Eq.~(12)
is worked out in the caption of Figure~1.  Although Eq.~(12) is 
derived here from a crude, heuristic argument, it is confirmed by
a rigorous analysis of systems that have a symbolic dynamics.$^{21}$
%Schack1996a

There are limits to the validity of Eq.~(12).  First, Eq.~(12) is
no longer valid for times large enough that the partition of the 
unperturbed pattern begins to have more than one piece in each 
coarse-grained cell, because then the perturbations of two such pieces 
are correlated.  Second, Eq.~(12) goes bad when the allowed entropy 
increase $\Delta H$ is sufficiently small---somewhere between 0 and 
$D$ bits---because one is then required to keep track of the perturbed 
pattern on scales finer than the width of the unperturbed pattern.  
The information $\Delta I$ then counts perturbation histories as 
distinct if the corresponding perturbed patterns differ only on scales 
finer than the finest scale set by the system dynamics.  As already 
indicated, for a diffusive perturbation $\Delta I$ becomes infinite 
as $\Delta H$ goes to zero because a diffusive perturbation has an 
infinite number of realizations.  The result is that when $\Delta H$ 
is sufficiently small, the information $\Delta I$ reveals properties 
of the stochastic perturbation---essentially the number of perturbation 
histories that differ on very fine scales.

The flip side of the coin is that for $\Delta H\gsim D$, the 
information-entropy relation~(12) is, like the entropy increase of
Eq.~(11), a property of the system dynamics, not a property of the 
perturbation.  This is evident from the fact that both the entropy
increase of Eq.~(11) and the information-entropy relation of Eq.~(12) 
are determined by the KS entropy and are independent of the strength 
of the perturbation, provided the perturbation is strong enough to 
become effective before the unperturbed pattern spreads over all the 
coarse-grained cells.    

The further entropy increase $\Delta H_0-\Delta H$ beyond the allowed
increase $\Delta H$ is a logarithmic measure of the number of 
{\it nonoverlapping\/} partial densities of volume 
$2^{-(\Delta H_0-\Delta H)}{\cal V}_0$ that fit within the average
density of volume ${\cal V}_0$.  In contrast, the information 
$\Delta I$ is a logarithmic measure of the much greater number of 
{\it overlapping\/} partial densities produced by the perturbation. 
The proliferation of overlapping partial densities is a consequence of 
the chaotic dynamics, which spreads the unperturbed pattern over an 
exponentially increasing number of coarse-grained cells, in each of 
which the perturbation acts essentially independently.

The notion of hypersensitivity to perturbation can be applied directly
to quantum dynamics.  The initial data specify a state vector that 
is localized on phase space.  This state vector evolves under the 
influence of an unperturbed dynamics and a stochastic perturbation.  
Each perturbation history leads to a particular state vector at time
$t$.  Averaging over the perturbation yields an average density operator 
that corresponds to an entropy increase $\Delta H_0$.  Computer simulations 
indicate that quantum systems whose classical limit is chaotic display 
hypersensitivity to perturbation,$^{22}$
%Schack1996b
in that the information $\Delta I$ about the perturbation required to 
reduce the entropy increase to $\Delta H$ far exceeds the entropy reduction
$\Delta H_0-\Delta H$.  

The mechanism for classical hypersensitivity is that the chaotic dynamics 
spreads the phase-space pattern over an exponentially increasing number 
of phase-space cells, in each of which the perturbation acts independently.  
The result is a proliferation of overlapping perturbed phase-space 
patterns.  Our simulations suggest a similar mechanism for quantum 
hypersensitivity.  The chaotic quantum dynamics spreads the state vector 
over an exponentially increasing number of quantum phase-space cells, 
each of which is a state vector localized on phase space.  Since the 
evolution is unitary Schr\"odinger evolution, the spreading creates a 
coherent superposition of these localized state vectors.  The stochastic 
perturbation changes amplitudes and phases within this superposition, 
thereby creating a proliferation of {\it nonorthogonal\/} state vectors, 
which are distributed randomly over the space spanned by the localized 
state vectors in the superposition.  This proliferation of nonorthogonal 
state vectors is responsible for quantum hypersensitivity.

Our computer simulations indicate that, in contrast to the classical 
situation, quantum hypersensitivity can occur for a perturbation that
is correlated across all of phase space.  The essential difference seems
to be that quantum mechanically the perturbation can act on the phases
in the quantum superposition, a mechanism not available to a classical
perturbation.  Should this speculation be correct, quantum hypersensitivity
would emerge as a distinctly quantum-mechanical phenomenon, similar to
classical hypersensitivity, yet subtly different because of quantum
superposition.  Indeed, one could say that in the case of classical 
hypersensitivity, the perturbation generates {\it classical information\/}
that is stored in an ensemble of overlapping phase-space patterns, 
whereas in the case of quantum hypersensitivity, the perturbation 
generates {\it quantum information\/} that is stored in an ensemble
of nonorthogonal state vectors.

Though our simulations are not sufficient to verify this mechanism
for quantum hypersensitivity, they do suggest models that might be 
simple enough to elucidate the nature of quantum hypersensitivity 
analytically.  Such models, together with further computer simulations, 
are the focus of our current work, the goal of which is to develop a 
deeper understanding of the unpredictability of open quantum systems.

\section{ACKNOWLEDGMENTS}
This work was supported in part by the Phillips Laboratory 
(Grant No.~F29601-95-0209) and by the National Science Foundation 
through its support of the Institute for Theoretical Physics at 
the University of California at Santa Barbara (Grant No.~PHY94-07194).

\vfill\eject

\line{}
\vglue 0.1truein

\centerline{\bf REFERENCES}

\medskip
\frenchspacing

\ref %Laplace1812
P.~S. Laplace. {\it Th\'eorie Analytique des Probabilit\'es}, Courcier, 
Paris, 1812.

\ref %Jeffreys1961
H.~Jeffreys. {\it Theory of Probability}, 3rd ed., Clarendon Press, Oxford, 
1961.

\ref %Jaynes1989b
E.~T. Jaynes. {\it Papers on Probability, Statistics and Statistical 
Physics}.  R.~D. Rosen\-krantz (Ed.),  Kluwer Academic, Dordrecht, 1989.

\ref %DeFinetti1990
B.~de~Finetti. {\it Theory of Probability}, Wiley, New York, 1990.

\ref %VonMises1921
R.~von Mises. {\"U}ber die gegenw\"artige Krise der Mechanik.
{\it Zeitschrift f\"ur angewandte Mathematik und Mechanik\/} {\bf 1}: 
1921, pp.~425--431.

\ref %Jaynes1986
E.~T. Jaynes.  Monkeys, kangaroos, and $N$. In {\it Maximum Entropy and
Bayesian Methods in Applied Statistics}.  J.~H. Justice (Ed.), Cambridge
University Press, Cambridge, England, 1986, pp.~26--58.

\ref %Savage1972
L.~J. Savage. {\it The Foundations of Statistics}, 2nd ed., Dover, New 
York, 1972.

\ref %Berger1985
J.~O. Berger. {\it Statistical Decision Theory and Bayesian Analysis}, 
2nd ed., Springer, Berlin, 1985.

\ref %Shannon1949
C.~E. Shannon and W. Weaver. {\it The Mathematical Theory of Communication},  
University of Illinois Press, Urbana, IL, 1949.

\ref %Gallager1968
R.~G. Gallager. {\it Information Theory and Reliable Communication}, 
Wiley, New York, 1968.

\ref %Cover1991
T.~M. Cover and J.~A. Thomas. {\it Elements of Information Theory},
Wiley, New York, 1991.

\ref %Peres1993
A.~Peres. {\it Quantum Theory: Concepts and Methods}, Kluwer Academic, 
Dordrecht, 1993, Part~II.

\ref %Bell1987
J.~S. Bell.  {\it Speakable and Unspeakable in Quantum Mechanics}, Cambridge
University Press, Cambridge, England, 1987.

\ref %Lichtenberg1983
A.~J. Lichtenberg and M.~A. Lieberman. {\it Regular and Chaotic Dynamics}, 
2nd ed., Springer, New York, 1992.

\ref %Alekseev1981
V.~M. Alekseev and M.~V. Yakobson.  Symbolic dynamics and hyperbolic
dynamic systems. {\it Phys. Reports\/} {\bf 75}: (1981), pp.~287--325.

\ref %Fox1990
R.~F. Fox.  Chaos, molecular fluctuations, and the correspondence limit.
{\it Phys. Rev.~A\/} {\bf 41}: 1990, pp.~2969--2976.

\ref %Caves1993b
C.~M. Caves.  Information, entropy, and chaos. In {\it Physical Origins
of Time Asymmetry}.  J.~J. Halliwell, J.~P\'erez-Mercader, and
W.~H. Zurek (Eds.), Cambridge University Press, Cambridge, England, 1994,
pp.~47--89.

\ref %Schack1992a
R.~Schack and C.~M. Caves. Information and entropy in the baker's map.
{\it Phys. Rev. Lett.} {\bf 69}: 1992, pp.~3413--3416.

\ref %Schack1993e
R.~Schack and C.~M. Caves. Hypersensitivity to perturbations in the
quantum baker's map. {\it Phys. Rev. Lett.} {\bf 71}: 1993, pp.~525--528.

\ref %Schack1994b
R.~Schack, G.~M. D'Ariano, and C.~M. Caves. Hypersensitivity to
perturbation in the quantum kicked top. {\it Phys. Rev.~E} {\bf 50}:
1994, pp.~972--987.

\ref %Schack1996a
R.~Schack and C.~M. Caves. Chaos for Liouville probability densities.
{\it Phys. Rev.~E\/} {\bf 53}: 1996, pp.~3387--3401.

\ref %Schack1996b
R.~Schack and C.~M. Caves. Information-theoretic characterization of quantum
chaos.  {\it Phys. Rev.~E\/} {\bf 53}: 1996, pp.~3257--3270.

\ref %Jaynes1957a
E.~T. Jaynes. Information theory and statistical mechanics. 
{\it Phys. Rev.} {\bf 106}: 1957, pp.~620--630.

\ref %Jaynes1957b
E.~T. Jaynes. Information theory and statistical mechanics.~II. 
{\it Phys. Rev.} {\bf 108}: 1957, pp.~171--190.

\ref %Landauer1961
R.~Landauer. Irreversibility and heat generation in the computing process. 
{\it IBM J. Res. Develop.} {\bf 5}: 1961, pp.~183--191.

\ref %Landauer1988
R.~Landauer. Dissipation and noise immunity in computation and 
communication. {\it Nature\/} {\bf 355}: 1988, pp.~779--784.

\ref %Bennett1982
C.~H. Bennett.  The thermodynamics of computation---a review. 
{\it Int. J. Theor. Phys.} {\bf 21}: (1982), pp.~905--940.

\ref %Zurek1987a
W.~H. Zurek. Thermodynamic cost of computation, algorithmic complexity 
and the information metric.  {\it Nature\/} {\bf 341}: 1989, pp.~119--124.

\ref %Zurek1989b
W.~H. Zurek. Algorithmic randomness and physical entropy.  
{\it Phys.\ Rev.\ A\/} {\bf 40}: 1989, pp.~4731--4751.

\ref %Caves1989
C.~M. Caves.  Entropy and information: How much information is needed
to assign a probability?  In {\it Complexity, Entropy, and the Physics
of Information}, Santa Fe Institute Studies in the Sciences of Complexity,
Proceedings Vol.~VIII.  W.~H. Zurek (Ed.), Addison-Wesley, Redwood City,
California, 1990, pp.~91--115.

\ref %Szilard1929
L.~Szilard. \"Uber die Entropieverminderung in einem thermodynamischen 
System bei Eingriffen intelligenter Wesen (On the decrease of entropy in 
a thermodynamic system by the intervention of intelligent beings).  
{\it Z.~Phys.} {\bf 53}: 1929, pp.~840--856.  English translation in 
H.~S. Leff and A.~F. Rex (Eds.).  {\it Maxwell's Demon: Entropy, 
Information, and Computing}, Adam Hilger, Bristol, 1990. 

\nonfrenchspacing

\vfil\eject

\line{}

\vskip 0.1truein plus 0.1fil

\centerline{\bf FIGURE CAPTION}

\medskip\nobreak

Figure~1. \ Cartoon of classically chaotic Hamiltonian dynamics.  
Phase space is represented as a two-dimensional square gridded into 
coarse-grained cells.  The initial data are that the system begins 
in the shaded coarse-grained cell shown in~(a); these initial data 
correspond to a uniform phase-space density over the shaded cell.  Under 
chaotic Hamiltonian evolution, the phase-space density spreads across 
phase space, creating a pattern of uniform density, shown as the central 
dark line in~(b); the evolved pattern occupies the same phase-space volume 
as the initial cell and develops structure on finer and finer scales as 
the evolution proceeds.  The chaotic dynamics is characterized by the 
{\it Kolmogorov-Sinai\/} (KS) or {\it metric entropy}, denoted by $K$.  
A crude picture is that the evolved pattern spreads over $2^{Kt}$ 
coarse-grained cells at time $t$.  To analyze hypersensitivity to 
perturbation, we assume that the evolution is perturbed stochastically 
by a diffusive perturbation that is essentially independent from one 
coarse-grained cell to the next.  A typical perturbed pattern is shown 
in~(b) as the dashed line that is twined about the unperturbed pattern.  
The average density, shown as the shaded region in~(b), is obtained by 
averaging over all the perturbed patterns.  In this example the average 
density occupies a phase-space volume that is about four times as large 
as the volume occupied by the unperturbed pattern, corresponding to an 
entropy increase of $\Delta H_0\sim\log_2 4=2$ bits.  To reduce the 
entropy increase to $\Delta H=1$ bit, one must answer the following 
question: in each coarse-grained cell, on which side of the unperturbed 
pattern does the perturbed pattern lie?  Answering this question requires 
giving $\Delta H_0-\Delta H\sim1$ bit of information {\it for each 
coarse-grained cell occupied by the unperturbed pattern\/} and thus 
requires a total amount of information given by
$$
\Delta I\sim 2^{Kt}(\Delta H_0-\Delta H)\sim 2^{Kt}\;{\rm bits.}
$$

\vfil\eject
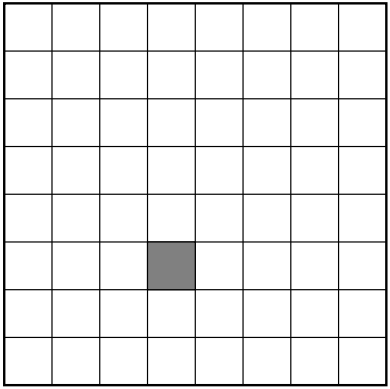
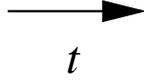
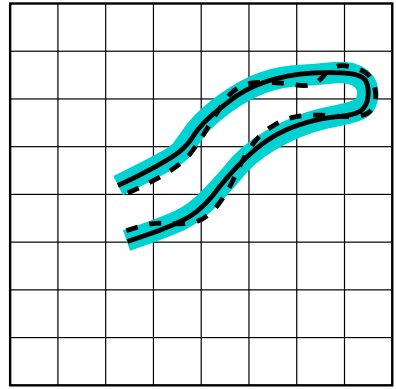
\end